# A multiple k-means cluster ensemble framework for clustering citation trajectories

Joyita Chakraborty[a], Dinesh K. Pradhan[b], Subrata Nandi[a]

[a]Department of CSE, National Institute of Technology, Durgapur, 713209, India
[b]Department of CSE/IT, Dr. B.C. Roy Engineering College, Durgapur, 713206, India

**Abstract**

Citation maturity time varies for different articles. However, the impact of all articles is measured in a fixed window (2-5 years). Clustering their citation trajectories helps understand the knowledge diffusion process and reveals that not all articles gain immediate success after publication. Moreover, clustering trajectories is necessary for paper impact recommendation algorithms. It is a challenging problem because citation time series exhibit significant variability due to non-linear and non-stationary characteristics. Prior works propose a set of arbitrary thresholds and a fixed rule-based approach. All methods are primarily parameter-dependent. Consequently, it leads to inconsistencies while defining similar trajectories and ambiguities regarding their specific number. Most studies only capture extreme trajectories. Thus, a generalized clustering framework is required. This paper proposes a *feature-based multiple k-means cluster ensemble framework*. Multiple learners are trained for evaluating the credibility of class labels, unlike single clustering algorithms. 1,95,783 and 41,732 well-cited articles from the Microsoft Academic Graph data are considered for clustering short-term (10-year) and long-term (30-year) trajectories, respectively. It has linear run-time. Four distinct trajectories are obtained– *Early Rise-Rapid Decline (ER-RD)* ( 2.2%), *Early Rise-Slow Decline (ER-SD)* ( 45%), *Delayed Rise-No Decline (DR-ND)* ( 53%), and *Delayed Rise-Slow Decline (DR-SD)* ( 0.8%). Individual trajectory differences for two different spans are studied. Most papers exhibit *ER-SD* and *DR-ND* patterns. The growth and decay times, cumulative citation distribution, and peak characteristics of individual trajectories' are re-defined empirically. A detailed comparative study reveals our proposed methodology can detect all distinct trajectory classes.

## 1. Introduction

A citation trajectory represents the time-series distribution of annual citations received by a paper [1]. The other terms used for a citation trajectory are 'citation curve,' 'citation pattern,' 'citation histories,' and 'citation time series.' Clustering citation trajectories refers to grouping papers with similar shapes or identical patterns in their citation life cycle [2]. It is a fundamental source of time and topic-correlated information in scholarly networks. Thus, it can capture the knowledge diffusion process [3].

Clustering trajectories can identify hidden patterns such as how some information gets early attention (pre-mature discovery), how some information receives attention for a short period and fades out soon, how some information gets attention and remains relevant indefinitely, how some information gets unnoticed and gets delayed attention (delayed recognition) [1], etc. Besides, it is crucial in paper impact recommendation algorithms for measuring similarity distances [4].

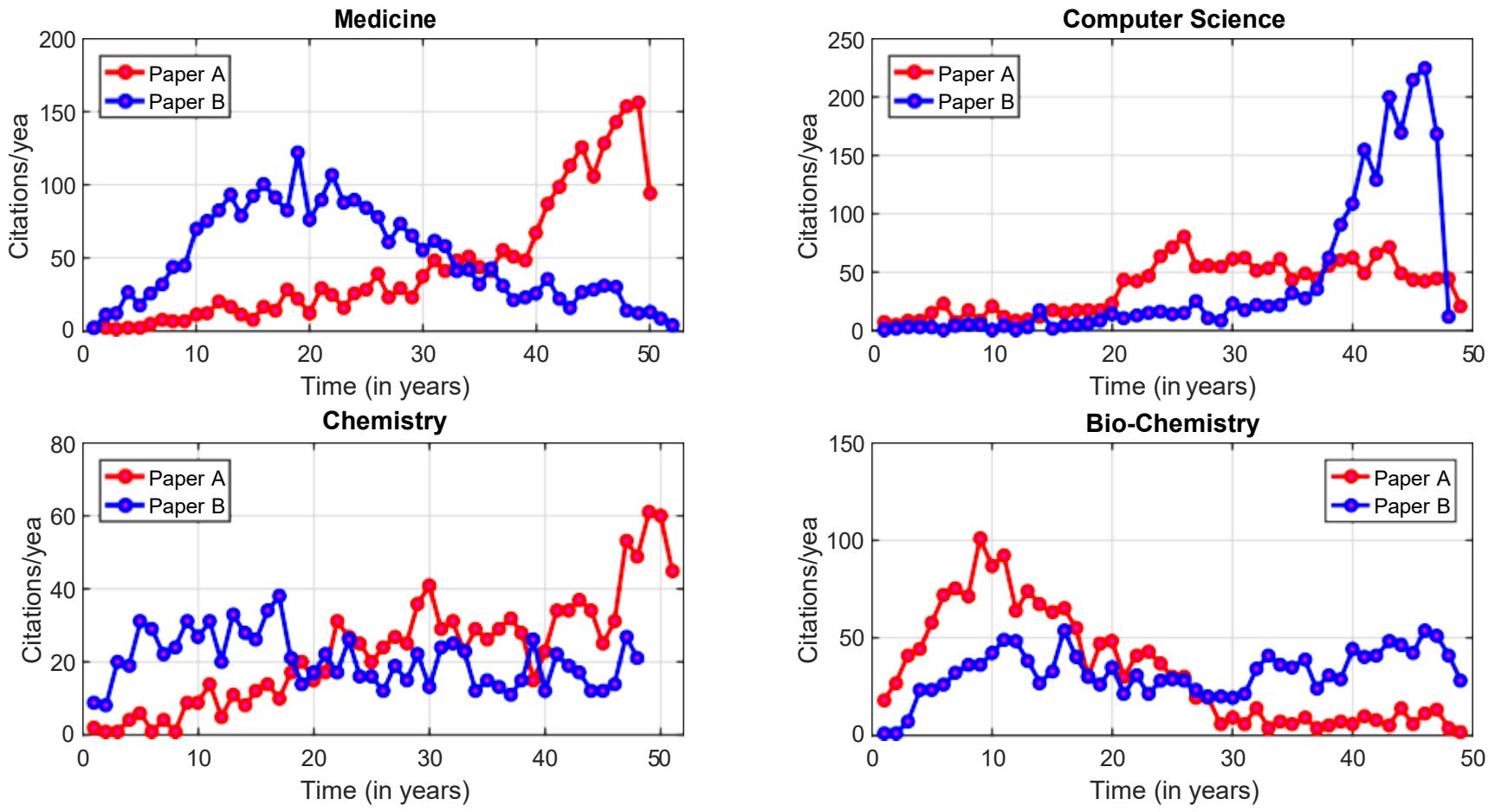

Figure 1: **Citation trajectories are plotted for two randomly chosen papers, each from four fields– Medicine, Computer Science, Chemistry, and Bio-Chemistry.**

Traditional research evaluation metrics calculate total or average citation counts accumulated within a specific time window [5, 6]. However, they can not truly capture the time-varying changes in a publication's impact. We present an example to illustrate this. The citation trajectories of two randomly chosen papers from four fields are plotted in figure 1. Both articles from each field in the Microsoft Academic Graph (MAG) dataset [7] were published around the same time and got equal citations. Papers from medicine, computer science, chemistry, and biochemistry fields received 2,593, 1,959, 1,166, and 1,607 citations in around 50 years time, respectively. Comparing early citations of two articles in the medicine field, we find that paper B is more likely to achieve a higher impact in the future than paper A. However, paper A suddenly jumps to receive greater than 50 citations annually after 40 years of publication. Although both papers receive the exact total and average citations, there is considerable variability in their temporal trajectories. Thus, different articles differ significantly in their citation maturity times. Clustering helps in empirically understanding the maturity cycles (growth and decay) of individual trajectory classes. It will also help in cross-discipline evaluation, where some fields may require more time to grow than others [8].

It is a challenging problem because citation trajectories exhibits non-linear and non-stationary properties when observed on a time scale [2, 9]. Further, exponential growth in bibliographic databases has added to this variance. Consequently, Zamani et al. [10] mathematically prove that citation trajectories diffuse anomalously, and their variance varies proportional to $t^{2H}$ where $H \neq 1/2$. Moreover, S. Baumgartner and L. Leydesdorff [11] report that a fifth-order polynomial fits citation trajectories of articles over 16 years. Thus, it is a more complex phenomenon than expected. Next, a comprehensive survey of existing literature is presented.

### *1.1. Background*

In sub-section 1.1, a detailed review of the clustering literature is discussed chronologically.

The clustering problem has been roughly studied since the 1980s. E. Garfield [12] first identified the *Delayed Recognition (DR)* phenomenon. Highly-cited papers were considered cited for over 10+ years with few initial citations. E. Aversa [13] clustered trajectories of 400 highly-cited articles over 9 years. They used a simple k-means on raw time series and detected two clusters– *Early Rise-Rapid Decline (ER-RD)* and *Delayed Rise-Slow Decline (DR-SD)*.

D. Aksnes [14] clustered 297 articles cited over 16 years. For each article, he calculated citations received in 3 and 7-12 years of time windows. The trajectory rise was categorized into *Early*, *Medium*, and *Delayed rise* if a publication got $> 30\%$, between 15%-30%, and $< 15\%$ of its final citations in the first 3 years, respectively. Further, the decline was categorized into *Rapid*, *Slow*, and *No Decline* if a publication received $< 30\%$, between 30%-50%, and $> 50\%$ of its final citations in the later period. Three clusters were identified– *ER-RD*, *MR-SD*, and *DR-ND*.

Raan et al. [15] in 2004, first coined the term *'Sleeping Beauty (SB)'* for the *DR* phenomenon. S. Redner [16] identified three clusters by arbitrarily defining thresholds– *Sleeping Beauties (SB), Discovery Papers (DP),* and *Hot Papers (HP)*. SB's received $> 250$ citations with the ratio of the mean citation age to publication age (r) $> 0.7$. DP's received $> 500$ citations with r $< 0.4$. HP's received $> 350$ citations with r $> 2/3$. Costas et al. [17] initially divided the trajectory into two halves based on the time taken to attain 50% of its total citations (Y50). Next, they determine the peak location by comparing the time of receiving 25% and 75% of its total citations with Y50. They considered articles cited over 29 years. They identified three clusters– *'Flashes-in-the-Pan (FP),' 'Normal Documents (ND),'* and *'Delayed Documents (DD).*

Li and Ye [18] identified a sub-category of SB's – All-Element-Sleeping- Beauties (ASB). Jian Li [19] separately identified three clusters – *ASB, FP,* and *DR*. Other studies proposed different thresholds for SBs– heartbeat spectra [20] and awakening intensity of SBs [21].

Few studies [11, 22] grouped them into two clusters- *transient* and *sticky* knowledge claim. Chakraborty et al. [23] considered the number of peaks ($n_{CP}$) and their location ($t_{CP}$) for defining thresholds. They defined six clusters– *PeakInit* ($n_{CP}$ in $t_{CP} <= 5$ years followed by an exponential decline), *MonInc* (monotonic increase in $n_{CP}$ till 20 years after publication), *PeakMult* (multiple $n_{CP}$), *MonDec* ($n_{CP}$ in the first year after publication followed by a monotonic decrease in citations), *PeakLate* (few initial citations and single $n_{CP}$ in $t_{CP} > 5$ years but not in the last year), and *Others* (undefined trajectory).

Bjork et al. [24] and Min et al. [25] used the BASS model from management studies. Min et al. [25] considered two parameters– *innovation (p)* and *imitation (q)* coefficients and defined four clusters – papers with *(small p, small q)*, *(large p, small q)*, *(small p, large q)*, and *(large p, large q)* values. G. Colavizza and M. Franceschet [26] proposed a shape-based approach and non-linear spectral-clustering method. Three clusters were identified – *sprinters*, *middle-of-the-roads*, and *marathoners*. Zhang et al. [27] proposed a model-based approach and a simple k-means algorithm. Four clusters were identified– *normal low, normal high, delayed documents,* and *evergreens.* Bornmann et al. [28] measured field and time normalized citation impact scores and identified two clusters– *Hot Papers (HP)* and *Delayed Recognition (DR)*. The thresholds were defined based on peaks in the early or later half period, similar to Costas et al. [17].

F. Ye and L. Bornmann [29] categorized into two clusters– *Smart Girls (SG)* and *Sleeping Beauties (SB)*. They used beauty co-efficient [30] and proposed the concept of citation angles. The citation angle for *SGs* was $> 60^o$ and $< 30^o$ for *SBs* as compared to the zero citation line. Besides, over the past decade, many works [30, 31, 32] have only studied *SBs* from multiple dimensions. He et al. [33] separately modeled SBs into *single-peak SBs*, *second-act SBs*, and *second-act non-SBs*. Recently, Gou et al. [34] defined papers receiving multiple citation peaks even after decay as exhibiting a *literature revival* phenomenon.

Summarizing, we observed that trajectories with similar shapes and inherent behavior are studied under different clusters. Clusters identified as– ER-RD [13, 14], FP [17], sprinters [26], transient-knowledge-claim [11], MonDec [23], HP [28], and SGs [29] represent similar trajectories. Besides, clusters identified as– DR [12, 28], SBs [16, 30, 31, 29], DD [27], and PeakLate [23] represent similar trajectories. Moreover, clusters identified as DR-ND [14], delayed rise [17], sticky-knowledge-claim [11], marathoners [26], evergreens [27], and MonInc [23] represent similar trajectories. Here, we briefly point out similarities between some of the clusters. A comprehensive comparison is made in section 3.4. Besides, most studies only capture two extreme trajectories– HP and SB. HP receive an initial citation burst followed by a rapid decay. SBs are papers with negligible initial citations followed by a citation burst in the later period. Thus, different methods capture different groups of trajectories. Consequently, there are ambiguities regarding the exact number of distinct trajectories.

The primary motivation draws from varying thresholds and arbitrary methods used for identifying similar trajectories [34]. Broadly, any combination of citation count, the time to receive such a value, citation peaks, or the time gap for receiving such peaks are fixed to set arbitrary thresholds. Moreover, clustering using raw time series can add noise [35]. It results in inconsistent citation and temporal characteristics noted for similar trajectories and contradictory cluster behavior. Some studies [17, 23, 28, 27] only explore growth attributes of a trajectory and do not study its decay in detail. Mostly, the growth phase is categorized into two periods–the sleeping phase and the recognition or awakening phase [34, 31]. Thus, there is variability in growth and decline times. Besides, most methods have parameter dependence adding to irregularities. Such techniques result in a major proportion of articles remaining unclustered. For instance, methods proposed by Chakraborty et al. [23] could not define trajectories for 45% articles.

Our prime objective is to propose a generalized clustering framework that can capture all probable distinct trajectories. For this, the challenge of evaluating the credibility of cluster labels in unsupervised learning algorithms needs to be addressed. Additionally, we aim to define a generic feature set that can capture the temporal evolution of any trajectory. It will help to empirically redefine the characteristics of individual trajectories by their subjective assessment.

The present study attempts to address multiple gaps and, in doing so, makes important contributions. First, the study proposes a feature-based multiple k-means cluster ensemble framework for clustering trajectories. The feature-based approach removes the necessity of manually defining thresholds for different trajectories. Besides, it reduces the dimensionality and complexity of raw time series. It also helps input an even length vector. Moreover, unsupervised ensemble learning accurately evaluates the credibility of class labels. Thus, ambiguities regarding the number of distinct trajectories are resolved. Second, trajectory differences are studied considering different time lengths. Third, the growth and decay times, citation, and peak characteristics of individual trajectory clusters are empirically re-defined. Fourth, this is the first study to examine diverse literature and point out that identical trajectories were studied as different clusters. Next, we compare all of them with the final cluster sets obtained in this study to validate our proposed methodology. This study's proposed generalized clustering framework can determine all possible distinct trajectories rather than only capturing extreme classes.

The paper is organized as follows. In section 2, we define the feature set, the clustering methodology, experimental settings, and the MAG data set in the brief. Section 3 contains the main clustering results, cluster characteristics, and comparative study for validation. Section 4 concludes the research and discusses limitations and future implications.

## 2. Materials and methods

This section initially presents the choice of feature set and then the multiple k-means cluster ensemble algorithm (MKMCE) used for clustering citation trajectories. Further, we discuss the experimental settings and briefly describe the Microsoft Academic Graph (MAG) data set.

Table 1: **The generic feature set for evaluating trajectories**

| Summary | No. | Features | Notation | Description |
|---|---|---|---|---|
| **Time-related features** | F1 | Initial time | $T_i$ | It is time a paper takes to attain the geometric mean of its final citations. |
| | F2 | Growth time | $T_g$ | It is the difference between the time a paper takes to receive its highest annual citation (maximum) and geometric mean. |
| | F3 | Decay time | $T_d$ | It is the difference between the time when a paper is last cited (publication age) and the time a paper takes to receive its highest annual citation. |
| **Citation-related features** | F4 | Citation gain in $T_i$ | $C_{T_i}$ | It is the ratio of cumulative citations received by a paper in time $T_i$ to its final citations. $C_{T_i}(c) = \frac{\sum_{t=0}^{T_i} c_t}{c}$ |
| | F5 | Citation gain in $T_g$ | $C_{T_g}$ | It is the ratio of cumulative citations received by a paper in time $T_g$ to its final citations. $C_{T_g}(c) = \frac{\sum_{t=0}^{T_g} c_t}{c}$ |
| | F6 | Citation gain in $T_d$ | $C_{T_d}$ | It is the ratio of cumulative citations received by a paper in time $T_d$ to its final citations. $C_{T_d}(c) = \frac{\sum_{t=0}^{T_d} c_t}{c}$ |
| **Peak count-related features** | F7 | Number of peaks of high-intensity | $n_{p^h}$ | The total number of citation values in the time series greater than or equal to $\mu_i + 3\sigma_i$ is calculated separately for two periods, $T_g$ and $T_d$. |
| | F8 | Number of peaks of medium-intensity | $n_{p^m}$ | The total number of citation values in the time series greater than or equal to $\mu_i + 2\sigma_i$ is calculated separately for two periods, $T_g$ and $T_d$. |
| | F9 | Number of peaks of low-intensity | $n_{p^l}$ | The total number of citation values in the time series greater than or equal to $\mu_i + \sigma_i$ is calculated separately for two periods, $T_g$ and $T_d$. |

### 2.1. *Feature selection*

This study extracts features from a paper's raw time-series and converts it to a lower-dimension space. The dimensionality reduction reduces memory requirements [35]. It speeds up the clustering process as the distance calculation in clustering algorithms using raw time series can be computationally expensive [36]. Moreover, some distance measures are sensitive to noise. Thus, clustering using raw data may group time series similar in noises than identical in their characteristic shape [35]. It also helps us to input an even-length vector. Table 1 represents a complete description of features and their notation.

Figure 2 represents a hypothetical trajectory. Broadly, we divide it for any given paper into two phases– *growth* and *decay* period. One of the pioneer studies modeling citation distribution is the WSB model proposed by Wang et al. [37]. They highlighted the importance of two features– *impact time* and *immediacy time* for studying a trajectory. Here, impact time is the time to achieve the geometric mean of the final citation count, and immediacy time is when a paper receives its maximum or highest annual citation. We use these two variables–impact and immediacy time to measure early growth or initial time and growth time, respectively. Both of them refer to the growth period.

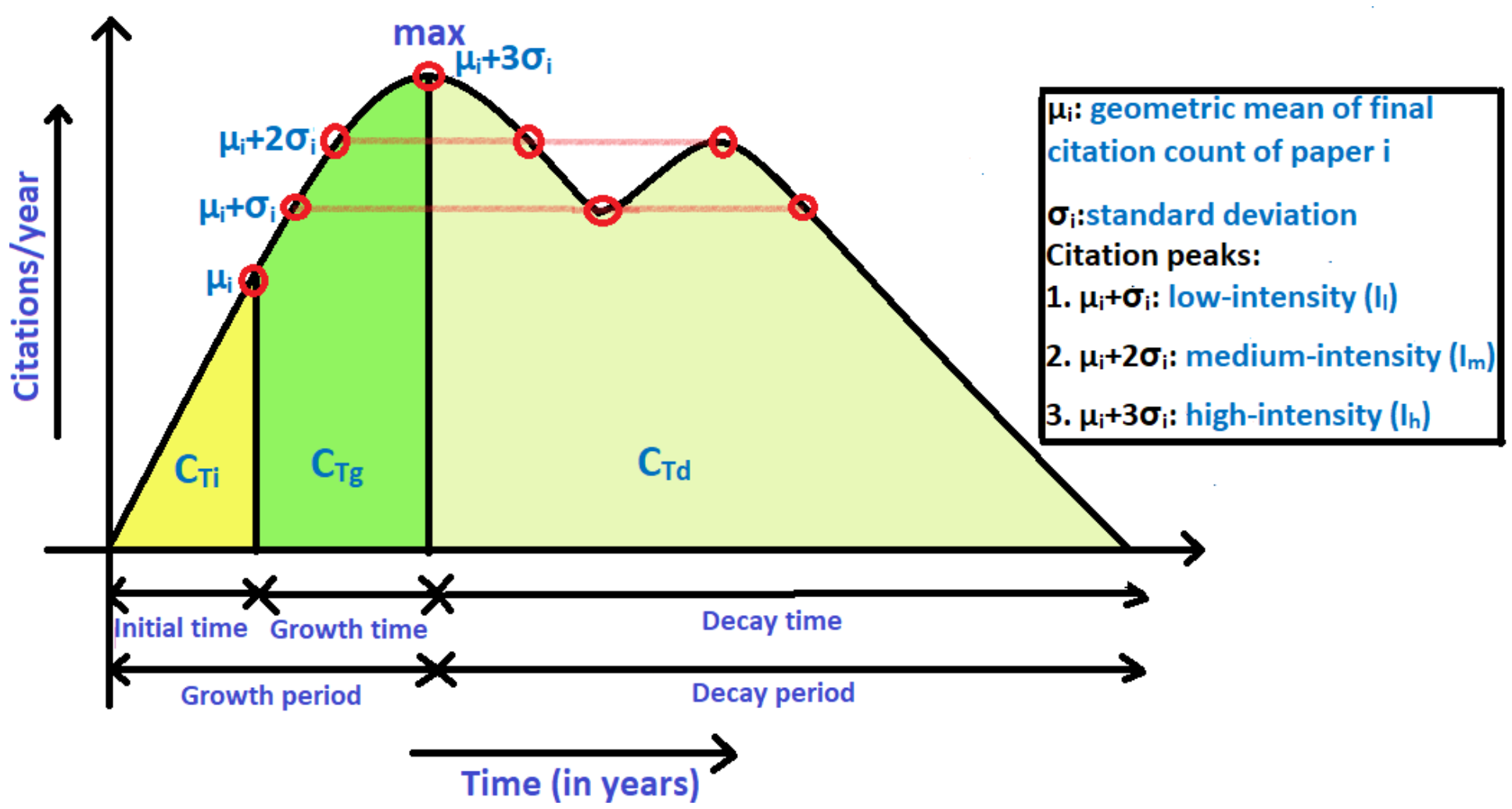

Figure 2: **A hypothetical citation trajectory of paper 'i' is plotted to understand the derived feature set quickly. Here, $\mu_i$ is the geometric mean of final citations, and $\sigma_i$ is the standard deviation considering the entire time series. The location of citation peaks is shown.**

Summarizing, we empirically define three measurements of time (time- related features F1, F2 and F3)– *initial time* ($T_i$), *growth time* ($T_g$), and *decay time* ($T_d$) (see full description in table 1). Further, we use statistical methods to measure a given paper's *citation growth or decay* (citation-related features F4, F5, and F6) separately in each of the above three times. We calculate the gain variable to measure citation growth or decline.

Finally, we separately measure *the number of peaks of three different intensities* (peak count-related features F7, F8, and F9) in growth and decay times. We use an empirical outlier detection rule [38, 36] widely used in time-series analysis. First, we calculate $i^{th}$ paper's mean citation ($\mu_i$) and standard deviation ($\sigma_i$) considering its entire time series. Next, we calculate citation values of magnitude greater than $\mu_i + \sigma_i$ (*low-intensity*), $\mu_i + 2\sigma_i$ (*medium-intensity*), and $\mu_i + 3\sigma_i$ (*high-intensity*). The resultant number of peaks of three different intensities separately occurring in growth and decay times are considered as input features (refer to figure 2).

It is essential to standardize features before clustering, as the distance between different data points is measured. Normalization refers to the probability distribution of features. We scale each feature ($f_i$) using a z-score value given as, $z_{f_i} = \frac{x_n - \mu_{f_i}}{\sigma_{f_i}}$

*2.2. Clustering methodology*

This section motivates the choice of multiple k-means cluster ensemble (MKMCE) algorithm [39] for clustering task. A cluster ensemble method is a more recent technique [40] to overcome the past limitations of single clustering algorithms.

*2.2.1. Problem definition*

Let X be an original data set consisting of N objects given by $\{x_i\}_{i=1}^{N}$. F is a set of M features. Besides, X(F) is the representation of X on F. It is a matrix of dimension N x M. Further, $X(f_j)$ represents $j^{th}$ column, $x_i(F)$ represents $i^{th}$ row, and $x_i(f_j)$ is the value of object $x_i$ in feature $f_j$.

1. *Multiple k-means clustering problem to generate a base cluster set:* The k-means algorithm is used as a base clusterer and is run multiple times to generate a base cluster set. Let, $Z = \{\zeta_h\}_{h=1}^{T}$ represent a set of T base clustering of X(F) where, $\zeta_h = \{C_{hl}\}_{l=1}^{kh}$ is the $h^{th}$ base clustering. $K = \{k_h\}_{h=1}^{T}$ represents the entire set of the number of clusters in each base clustering $\zeta_h$ where $k_h$ is the number of clusters in $h^{th}$ base clustering. Further, $\zeta_h(x_i(F))$ is the cluster label of object $x_i(F)$ in clustering $\zeta_h$. $\zeta_h(x_i(F)) = l$ denotes that object $x_i(F)$ belongs to cluster $C_{hl}$. The objective function (A) of k-means can be given as,

$$A(\zeta_h, y_h) = \sum_{l=1}^{k_h} \sum_{\zeta_h(x_i(F))=l,\, x_i(F)\in X(F)} d(x_i(F), y_{hl})^2 \quad (1)$$

Here, $y_{hl}$ is the $l^{th}$ cluster centre and $y_h = \{y_{hl}\}_{l=1}^{kh}$. Further, $d(x_i(F), y_{hl}) = \sqrt{|(x_i(F), y_{hl})|^2}$ is the Euclidean distance between object $x_i(F)$ and the centre $y_{hl}$ of the $l^{th}$ cluster. The algorithm minimizes A by constantly updating $\zeta_h$ and $y_h$. The clustering results are different each time due to different initial cluster centers. The next part resolves this issue.

2. *Cluster ensemble problem to generate a final cluster set:* The cluster ensemble problem integrates the base clusters rapidly to generate a final clustering $Z*$ of data set X(F) based on the clustering set Z. The final cluster set can be given as $C_*$. It can also be represented as $C_*=X(\zeta_*)$ where $\zeta_*$ is the final clustering feature. If $x_i(F) \in C_{*l}$ then, $x_i(F)(\zeta_*)=\zeta_{*l}$. Here, $\zeta_{*l}$ is the cluster label of $C_{*l}$ for $1 \leq i \leq N$.

*2.2.2. MKMCE algorithm*

It is an unsupervised ensemble learning method where several base clusters integrate to form final clusters with improved stability and accuracy. A cluster ensemble method is more efficient than other algorithms as it assesses the credibility of class labels. For instance, simple k-means is a linear clustering algorithm widely known for its low computational cost. However, it is susceptible to data distributions [41]. Besides, non-linear clustering algorithms, such as spectral clustering [26] and density-based spatial clustering of applications with noise (DBSCAN) [17], have expensive time costs. Their pair-wise distance calculation is not suitable for large data sets. Compared to them, Bai et al. [39] proved that the MKMCE algorithm is faster and can rapidly discover non-linearly separable clusters. It also worked well with large data sets such as KDD99, which had 1,048,576 entities.

Broadly, the MKMCE algorithm can be divided into two parts– (1) initially, it produces a base cluster set using k-means for multiple clusterings, (2) finally, integrate them into a final cluster set using cluster ensemble algorithm. The key steps are as follows:

1. *Local credibility function:* Unlike supervised learning algorithms, we do not know the exact cluster label of data points in unsupervised learning methods. Besides, a cluster center is used to represent a cluster in k-means. However, the cluster center is not convenient for representing a non-linear cluster. Therefore, a local credibility function is introduced to check whether the cluster label of objects falls into the local $\psi$ neighborhood of its cluster center. It is defined as,

$$\psi_h(x_i(F)) = \begin{cases} 1 & \text{if } x_i(F) \in \mathbf{N}(y_{hl}), \\ 0 & other\ wise, \end{cases} \tag{2}$$

Here, $l = \zeta_h(x_i(F))$ and $\mathbf{N}(y_{hl})$ is the $\psi$ neighborhood of the cluster center $y_{hl}$. It is also called as the local credible space of cluster $C_{hl}$, for $1 \leq i \leq N$ and $1 \leq h \leq T$.

2. *Multiple k-means clustering algorithm:* The objective function (U) to produce multiple base clusters using k-means can be given as,

$$\min_{Z} U(Z) = \sum_{h=1}^{T} \sum_{i=1}^{N} \theta_h(x_i(F))\, \psi_h(x_i(F))\, d(x_i(F), v_{h\zeta_h(x_i(F))}) \tag{3}$$

   Here, $\theta_h(x_i(F))$ is a boolean variable whose value is 1, if $x_i(F)$ takes part in $h^{th}$ base clustering and 0 otherwise. Also, once an object forms a base cluster, it does not participate in further k-means clustering. Thus, the aim is to minimize the value of objective function (U) with the constraint

$$\sum_{h=1}^{T} \theta_h(x_i(F))\, \psi_h(x_i(F)) = 1, 1 \leq i \leq N \tag{4}$$

   Initially, we set h=1, S=X(F), and $\theta_h(x_i(F))$=1 for $1 \leq i \leq N$. Next, we randomly select $k_h$ points as initial cluster centers from S. We apply the k-means algorithm multiple times with constraint equation 4. Only those objects are included in the cluster in the $\epsilon$ neighborhood of randomly initialized cluster centers. Finally, we assign each object in X(F)-S to the cluster with the nearest cluster centers. Next, we update $S = S \setminus S'$ where S' contains objects already clustered in $h^{th}$ base clustering. Also, update h=h+1 and $\theta_h(x_i(F)) = 1$, if $x_i(F) \in S$ else 0. This incremental clustering method runs until the desired number of base clusterings ($T_{max}$) is obtained, or the number of objects in $S'$ ($|S'|$) is lesser than $k_h^2$. $T_{max}$ can be set depending on user requirements. We initially set $T_{max}$ = 100, and based upon running it for a few iterations, we find $T_{max}$ = 10 as the ideal choice. Also, considering all prior works [23], we observed that up to a maximum of six clusters had been reported. The output of this part of the algorithm are base cluster set $Z = z_h$, $1 \leq h \leq T$ and a cluster center set $P = p_h$, $1 \leq h \leq T$.

3. *Cluster ensemble algorithm:* The final clusters should have high concurrence with the features of the base and original clusters. The overlap of credible local space between any two base clusters is used to measure their similarity. However, the credible local space between two base clusters is naturally small. It is due to the generation mechanism of base clusters using multiple k-means clustering. Thus, a latent cluster is introduced to measure the indirect overlap between base clusters.
Let, $C_{hl}$ and $C_{gj}$ be any two base clusters and $C_q$ be a latent cluster as per assumption. Let $y_{hl}$ and $y_{gj}$ represent the cluster centers of two base clusters. The center of the latent cluster can measure the similarity between any two base clusters. The mid-point of two cluster centers of base clusters is used to calculate the center of the latent cluster. It determines whether there is an indirect overlap between two base clusters. Further, it is calculated as, $d(y_{hl}, y_{gj}) = \frac{y_{hl}+y_{gj}}{2}$. If $d(y_{hl}, y_{gj}) \leq 4\epsilon$, the clusters are indirectly overlapped. Thus, the similarity measure is inversely proportional to $d(y_{hl}, y_{gj})$. Mathematically, it can be represented as,

$$\lambda(C_{h}, C_{g}) = \begin{cases} \frac{|P(\frac{y_{hl}, y_{gj}}{2})|}{d(y_{hl}, y_{gj})} & \text{if } d(y_{hl}, y_{gj}) \leq 4\epsilon, \\ 0 & otherwise, \end{cases} \quad (5$$

Finally, based on similarity measures, an undirected weighted graph is constructed G=(B, W). Here, B is a set of vertices representing a cluster label from Z, and W is the weight set. The similarity between any two clusters say, $C_x$ and $C_y$, is the weight of the edges between them. After constructing a weighted graph, a normalized graph cuts problem is proposed to derive cluster relations [42]. The larger similarity value between vertices represents that they belong to the same cluster with a higher probability. Also, they are dissimilar from the vertices of other clusters. Finally, the cluster labels are re-labeled to form the final clusters. Let $L(C_x)$ be the label of the subset which $C_x$ belongs to,

$$L(C_x) = l\,,$$

if $C_x \in A_l$, for $1 \leq l \leq k$ and $x \in A$. After re-labeling the base cluster set, Z can be transformed into a re-labeled set V as follows:

$$V_{x_i} = L(C_{hZ_h(x_i)})$$

for $1 \leq i \leq N$ and $1 \leq h \leq T$.

### *2.3. Experimental setup*

Let 'x' be a publication, and T = (1, 2, 3, ..., n) represent a series of consecutive years when 'x' is cited where $n \geq 1$ is an integer. For every $y \in T$, let $k_{x,y}$ represent the number of citations received by a paper x during a period y. A vector can define the citation trajectory of a publication x over T as:

$$k_{x,T} = (k_{x,1}, k_{x,2}, k_{x,3}, \ldots k_{x,n}) \tag{6}$$

The first fundamental question is, how many minimum citations should a paper receive to produce meaningful patterns in its trajectory? We use the metric *relative success ratio ($s_x$) of a paper x* as defined by Radicchi and Castellano [43]. It is given as,

$$s_x = \frac{c_x}{max(\mu_x, 5)} \tag{7}$$

Here, $c_x$ and $\mu_x$ are the total number of citations and mean citation rate received by a publication x over a period T. They can be represented as,

$$c_x = \sum_{T=1}^{n} k_{x,T} \tag{8}$$

$$\mu_x = \frac{c_x}{T} \tag{9}$$

G. Colavizza and M. Franceschet [26] use $max(\mu_x, 5)$ parameter in the denominator to account for years that receive a low mean citation rate. We only consider papers in our study with a value of $s_x \geq 1$. This method is used in the beginning to filter out well-cited articles with above-average citation impact so that its time-series trajectory can help us derive meaningful trends. It helps to address the problem of citation variance that occurs with various factors added due to the time of publication.

The following question is, what should be the exact length of a citation trajectory to be considered for clustering? Citation histories can only be compared if their lengths are equal [26]. Consequently, trajectories of three different time windows– 10 years, 20 years, and 30 years are considered separately. Above it, a minimum citation threshold, as discussed earlier, is set to filter out articles. The 10-year time window is chosen because impact factors [44] usually consider 2 to 5 years as it is the average time for attaining citation growth in most disciplines. However, some fields, such as social sciences, take longer to peak. Besides, we examine another 5-year window for capturing the decay pattern. Next, 20 and 30 years are multiples of 10, and it should allow us to investigate the long-term behavior of citation histories. Besides, the internet era began around 1985, and citation-based metrics such as the h-index came into practical use in 2005. Consequently, papers published in 1985, 1995, and 2005 and cited till 2015 are considered for studying 10-year [23, 26], 20-year [23], and 30-year trajectories [27], respectively.

### *2.4. Data*

The MAG data set [7] is used for empirical study in this paper. We get 1,95,783 papers published in 2005, 56,380 papers published in 1995, and 41,732 papers published in 1985 after filtering using methods described in section 2.3. All sets of papers receive citations till 2015. The cumulative citation distribution of papers is right-skewed. For example, while studying the 30-year distribution of cumulative citation count, we find that only 30 papers receive citations greater than 10,000, 106 papers receive citations greater than 5,000, 1,810 papers receive citations greater than 500, and 73.6% receive citations fewer than 100. Similarly, while studying the 10-year distribution of cumulative citation count, we find that only 28 papers receive citations greater than 10,000, 154 papers receive citations greater than 5,000, 5,362 papers receive citations greater than 500, and approximately 75% receive citations fewer than 100.

## 3. Results

This section extracts features and applies the multiple k-means cluster ensemble algorithm (MKMCE). The flowchart in figure 3 shows the number of papers obtained in each base and final cluster set. We do not find any significantly different clusters while considering data of the first two windows– 10 years and 20 years. Thus, the resulting distinct clusters are discussed considering two lengths– short-term (10 years) and long-term (30 years).

Table 2 represents the final cluster set. Further, the characteristics of each cluster are discussed, that is, citation trajectories growth and decay cycles. Finally, a comparative study is performed to validate obtained clusters with identical trajectories detected in prior literature.

Table 2: **Universal final cluster set**

| | **No Decline (ND)** | **Rapid Decline (RD)** | **Slow Decline (SD)** |
|---|---|---|---|
| **Early Rise (ER)** | - | Cluster S3 | Cluster L3, Cluster S2 |
| **Delayed Rise (DR)** | Cluster L1, Cluster S1 | - | Cluster L2 |

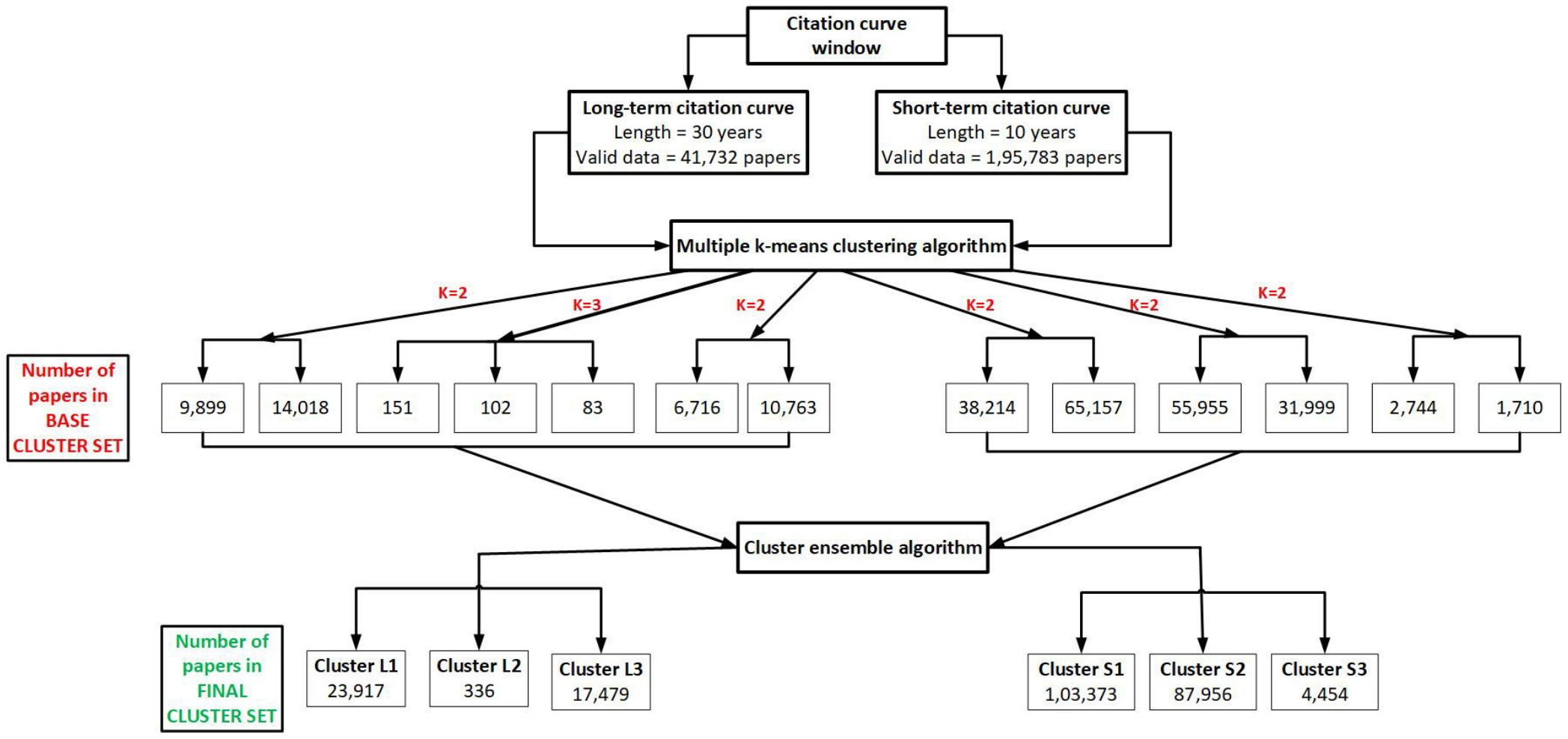

Figure 3: **The number of papers in each base and final cluster set after applying multiple k-means cluster ensemble algorithm.**

### *3.1. Clustering short-term trajectories*

The length of a citation trajectory considered for this study is 10 years. We consider papers published in the year 2005 and cited till 2015. The final set of papers considered is 1,95,783. We obtain three distinct clusters– cluster S1, S2, and S3 (see figure 3). The citation patterns for clusters S3, S2, and S1 are *'Early Rise-Rapid Decline (ER-RD)'*, *'Early Rise-Slow Decline (ER-SD)'*, and *'Delayed Rise-No Decline (DR-ND)'*, respectively (see table 2). Moreover, the percentage of papers in *ER-RD*, *ER-SD*, and *DR-ND* clusters are 2.2%, 45%, and 53%, respectively.

#### *3.1.1. Cluster analysis*

Table 3 shows each cluster's descriptive statistics of initial time, growth time, and decay time. Besides, the histogram plot in figure 4 shows cluster-wise cumulative citation distribution separately for three consecutive times. Finally, in figure 5, the box plot shows the distribution of citation peaks of different intensities.

1. *ER-RD:* They have an average initial time of 1 to 1.5 years and a growth time of $\sim$ 2 years. Thus, the total growth period for papers in this cluster is 3 to 3.5 years, followed by a quick decay in 2 years (see table 3). Cumulative citation distribution in figure 4 (a) reveals that they receive the least citations compared to other clusters as they receive citations only for a short window. However, a handful of papers also receive as high as 1,000 citations. M. Haghighat [45] points out self-citation stacking as one of the reasons behind the citation patterns of such hot papers. The number of $I^l$ intensity peaks are seen chiefly during their growth period (see blue colored box-plot in figure 5 (a)). Table 2 reveals it is only a characteristic of short-term citation trajectories.

Table 3: **(Short-term trajectory study) The descriptive statistics of initial time ($T_i$), growth time ($T_g$), and decay time ($T_d$) is shown for final clusters.**

| | Early Rise-Rapid Decline | | | Early Rise-Slow Decline | | | Delayed Rise-No Decline | | |
|---|---|---|---|---|---|---|---|---|---|
| **Short-term citation trajectory** | $T_i$ (in yrs.) | $T_g$ (in yrs.) | $T_d$ (in yrs.) | $T_i$ (in yrs.) | $T_g$ (in yrs.) | $T_d$ (in yrs.) | $T_i$ (in yrs.) | $T_g$ (in yrs.) | $T_d$ (in yrs.) |
| Mean | 1.51 | 2.16 | 2.11 | 2.31 | 3.15 | 3.88 | 3.05 | 5.72 | 0.5 |
| Standard deviation | 1 | 1 | 1 | 0.7 | 0.5 | 1 | 0.5 | 2 | 0.5 |
| Quartile 1 | 0 | 1 | 1 | 1 | 2 | 2 | 2 | 3 | 0 |
| Quartile 2 | 1 | 2 | 2 | 2 | 3 | 3 | 3 | 5 | 0 |
| Quartile 3 | 2 | 3 | 3 | 2 | 3 | 4 | 3 | 6 | 0 |

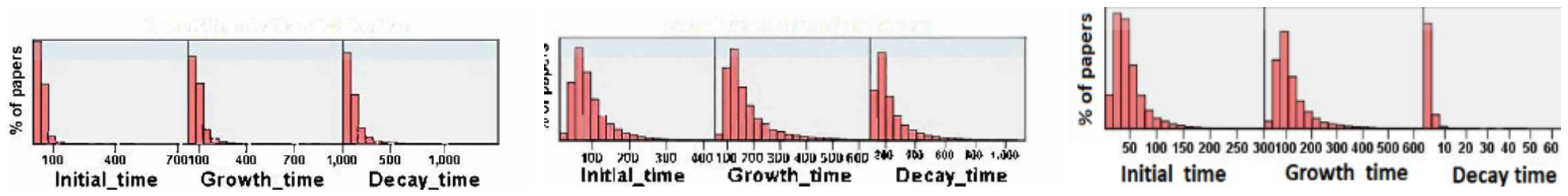

(a) Early Rise-Rapid Decline (b) Early Rise-Slow Decline (c) Delayed Rise-No Decline

Figure 4: **The histogram plot shows the cumulative citation distribution of each cluster for three consecutive times– initial time, growth time, and decay time. Figures (a), (b), and (c) represents ER-RD, ER-SD, and DR-ND clusters for short-term trajectories, respectively.**

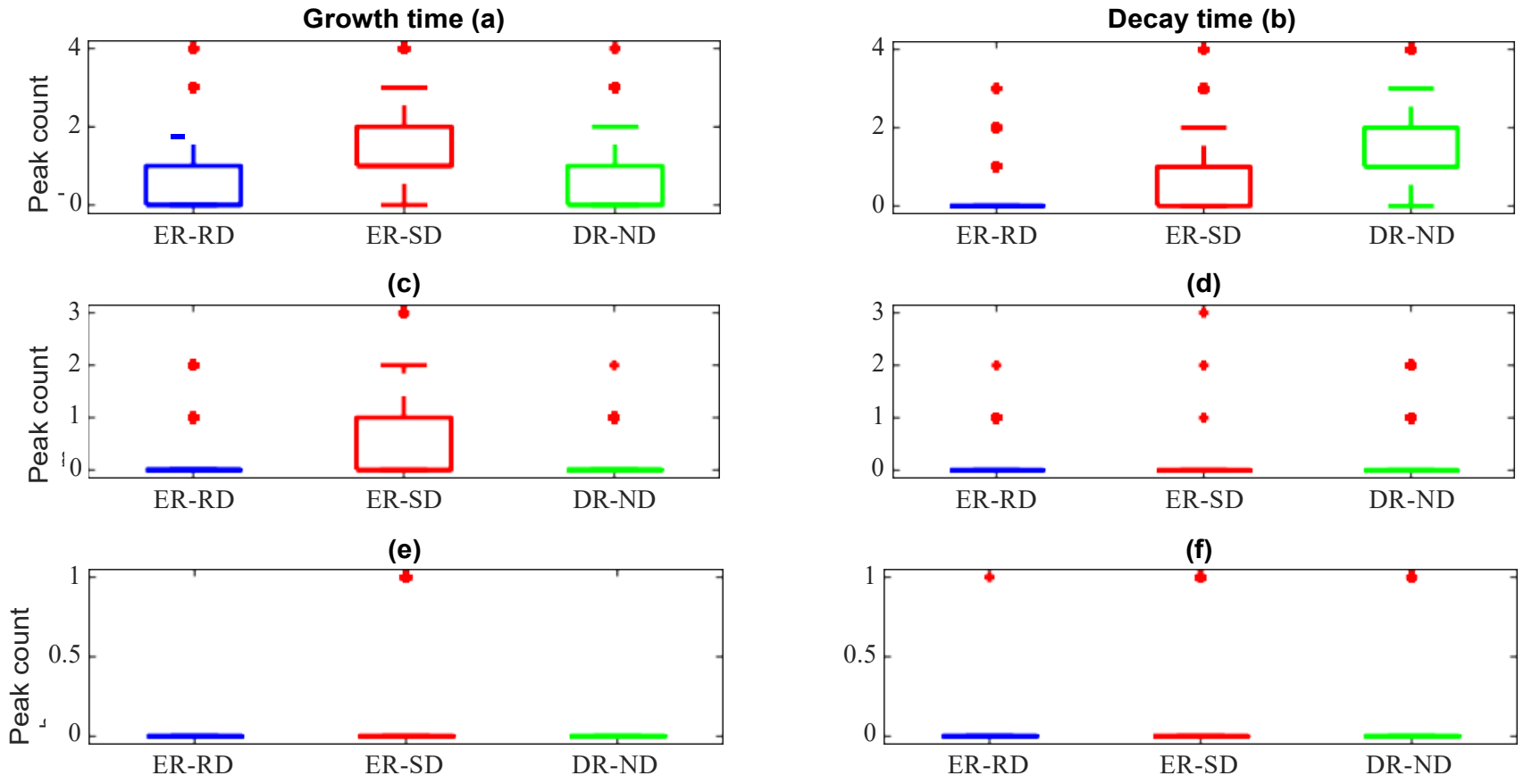

Figure 5: **The box plot represents the distribution of the number of citation peaks of three different intensities– $n_{I^l}$ (figures (a), (b)), $n_{I^m}$ (figures (c), (d)), and $n_{I^h}$ (figures (e), (f)) separately for growth and decay times. Blue, red, and green indicate ER-RD, ER-SD, and DR-ND clusters.**

2. *ER-SD:* They have an average initial time of 2 years and a growth time of 3 years. Consequently, the total growth time for papers in this cluster is 5 years, followed by a slow decay in the next 3 to 4 years (see table 3). The histogram plot depicting cumulative citation distribution in figure 4 (b) reveals a right-skewed distribution where papers receive more citations as it shifts from initial to growth to decay time. They receive the highest number of peaks of $I^l$ and $I^m$ intensity mainly in the growth time (see red colored box plot in figure 5 (a), (c)). Also, considering three consecutive times, they receive the maximum number of peaks compared to the other two clusters.
3. *DR-ND:* They have an average initial time of 3 years and an average growth time of $\sim$ 6 years. We observed no citation decay for the period analyzed. The histogram plot depicting cumulative citation distribution in figure 4 (c) shows that a significant proportion of citations is received in the growth time. They attain the highest number of citation peaks of $I^l$ intensity during their delayed growth time (see green colored box plot in figure 5 (b)).

### 3.2. *Clustering long-term trajectories*

The length of a citation trajectory considered for this study is 30 years. We consider papers published in the year 1985 and cited till 2015. The final set of papers considered is 41,732. We obtain three distinct clusters– cluster L1, L2, and L3 (see figure 3). The citation patterns for clusters L3, L1, and L2 are *'Early Rise-Slow Decline (ER-SD)'*, *'Delayed Rise-No Decline (DR-ND)'* and *'Delayed Rise-Slow Decline (DR-SD)'*, respectively (see table 2). The percentage of papers in *ER-SD*, *DR-ND*, and *DR-SD* clusters are 42%, 57%, and 0.8%, respectively.

#### 3.2.1. *Cluster analysis*

Table 4 shows each cluster's descriptive statistics of initial time, growth time, and decay time. Besides, the histogram bar plot in figure 6 shows cluster-wise cumulative citation distribution separately for three consecutive times. Finally, in figure 7, the box plot shows the distribution of citation peaks of different intensities.

1. *ER-SD:* They have an average initial time of 2 years and a growth time of 4 to 4.5 years. Consequently, the total growth time for papers in this cluster is 6 years, followed by an average decay time of 20 years (see table 4). Cumulative citation distribution in figure 6 (a) reveals ER-SD papers receive a significant proportion of their final citations during growth time.

Table 4: **(Long-term trajectory study) The descriptive statistics of initial time ($T_i$), growth time ($T_g$), and decay time ($T_d$) is shown for final clusters.**

| | Early Rise-Slow Decline | | | Delayed Rise-No Decline | | | Delayed Rise-Slow Decline | | |
|---|---|---|---|---|---|---|---|---|---|
| **Long-term citation trajectory** | $T_i$ (in yrs.) | $T_g$ (in yrs.) | $T_d$ (in yrs.) | $T_i$ (in yrs.) | $T_g$ (in yrs.) | $T_d$ (in yrs.) | $T_i$ (in yrs.) | $T_g$ (in yrs.) | $T_d$ (in yrs.) |
| Mean | 2.14 | 4.41 | 20.82 | 4.06 | 25.46 | 0 | 4.73 | 16.06 | 7.84 |
| Standard deviation | 1.22 | 2.85 | 3.72 | 2.00 | 2.93 | 0 | 2.42 | 2.63 | 2.09 |
| Quartile 1 | 1 | 3 | 18 | 2 | 22 | 0 | 2 | 6 | 5 |
| Quartile 2 | 2 | 4 | 20 | 4 | 25 | 0 | 4 | 16 | 7 |
| Quartile 3 | 3 | 6 | 22 | 4 | 26 | 0 | 5 | 17 | 7 |

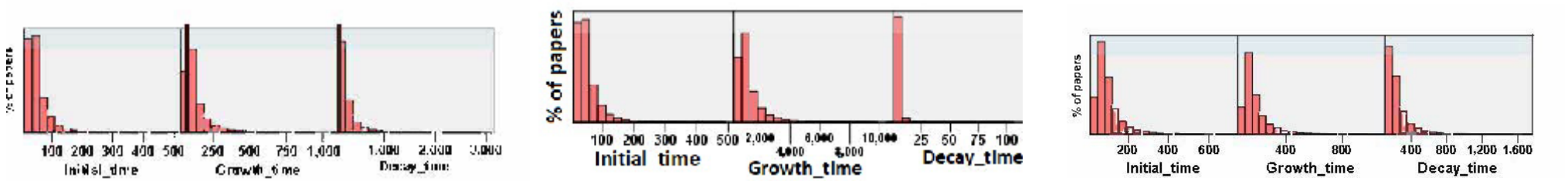

(a) Early Rise-Slow Decline (b) Delayed Rise-No Decline (c) Delayed Rise-Slow Decline

Figure 6: **The histogram plot shows the cumulative citation distribution of each cluster for three consecutive times– initial growth time, growth time, and decay time. Figures (a), (b), and (c) represents ER-SD, DR-ND, and DR-SD clusters for long-term trajectories, respectively.**

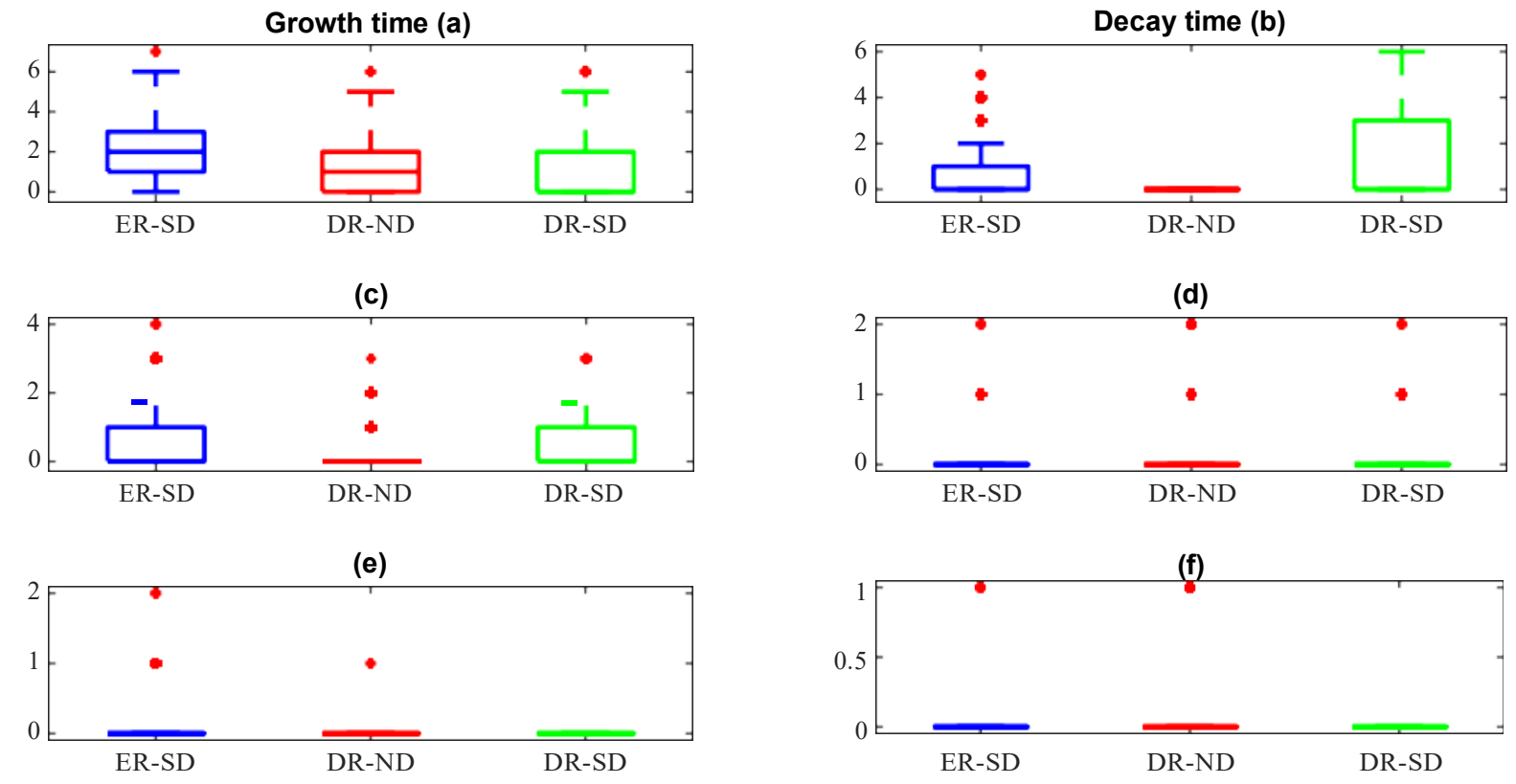

Figure 7: **The box plot represents the distribution of the number of citation peaks of three different intensities– $n_{Il}$ (figures (a), (b)), $n_{Im}$ (figures (c), (d)), and $n_{Ih}$ (figures (e), (f)) separately for growth and decay times. Blue, red, and green indicate ER-SD, DR-ND, and DR-SD clusters.**

They receive the highest number of $I^l$ and $I^m$ intensity peaks in growth time (see blue colored box plot in figure 7 (a) and (c)). Specifically, it gets a median of two peaks of $I^l$ intensity in growth time (see blue colored box plot in figure 7 (a)).

2. *DR-ND:* They have an average initial time of 4 years and an average growth time of 25 years. No significant decay is seen for the period analyzed (see table 4). Cumulative citation distribution in figure 6 (b) shows that they receive significant citations during their delayed growth period. Further, we find several $I_l$ intensity peaks during growth time (see red colored box plot in figure 7 (a)). It receives a single median peak and up to 5 peaks in the growth period.
3. *DR-SD:* They have an average initial time of 5 years and an average growth time of 16 years. Consequently, the total growth time for papers in this cluster is 16 to 20 years, followed by an average slow decay in the next 14 years (see table 4). Cumulative citation distribution in figure 6 (c) reveals that they receive negligible citations in the initial time, moderate citations in growth time, and a maximum proportion of citations in decay time. Low-intensity $I^l$ citation peak values are prominently visible in decay time (see green colored box plot in figure 7 (b)). Table 2 reveals it is only a characteristic of long-term citation trajectories.

### 3.3. Statistical cluster validation

The analysis of variance (ANOVA) test is conducted to validate the final clusters statistically. (Tables 1 and 2) [1] represents the ANOVA test results for short and long-term trajectories, respectively. We compare the mean values of a feature from k-groups and check whether the difference is statistically significant. It separates the variance into two components due to– mean differences and random influences [46]. We find that the final clusters in both studies are statistically significant (p-value<0.05), considering each feature separately. The largest F-values depicting feature importance are obtained for time-related features –*decay* and *growth time.*

### 3.4. Comparative study

This sub-section presents a qualitative comparison study between clusters to validate the proposed methodology. Table 5 and 6 examine how final cluster sets obtained in this study get mapped to identical clusters defined in prior literature. Here, a quantitative comparison is not feasible as different methods have different thresholds and parameter dependencies.

[1] https://github.com/decodejoyita/Clustering-citation-trajectories/tree/main

Table 5: **The ER-RD and DR-SD clusters are aligned with identical trajectories in literature**

| Final clusters | Brief description | Identical clusters in prior literature |
|---|---|---|
| Early Rise-Rapid Decline<br>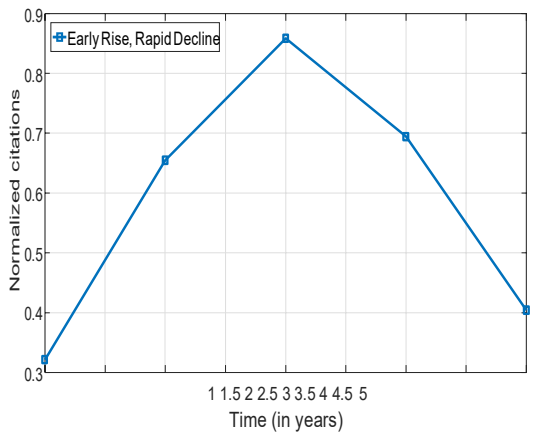<br> | **Short-term trajectory**<br>• *Growth time:* 3 years<br>• *Decay time:* 2 years<br>• *No. of papers:* 4,454 | • Early rise, rapid decline [13, 14]<br>• Flashes-in-the-pan [17]<br>• Sprinters [26]<br>• Transient-knowledge-claims [11]<br>• MonDec [23]<br>• Hot papers [28]<br>• Smart girls [29] |
| Delayed Rise-Slow Decline<br>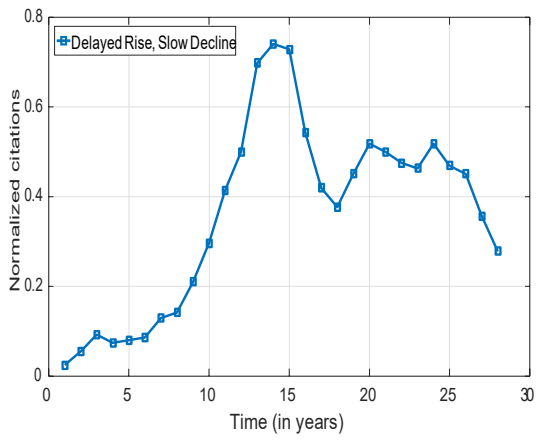<br> | **Long-term trajectory**<br>• *Growth time:* 16 years<br>• *Decay time:* 14 years<br>• *No. of papers:* 336 | • Revived classics [16]<br>• PeakLate [23]<br>• Sleeping beauties [15, 30, 31, 29]<br>• Delayed documents [27] |

E.Aversa [13] identified two clusters – 'Early Rise-Rapid Decline (ER-RD)' and 'Delayed Rise-Slow Decline (DR-SD).' ER-RD are defined as papers with a growth time of 3 years followed by a rapid decay. The exact decay time is not mentioned. DR-SD are defined as papers with a growth time of 6 years followed by a decline in the next 4 years. The ER-RD and DR-SD clusters are identical and align with *ER-RD* and *ER-SD* clusters obtained in this study.

D. Aksnes [14] identified three clusters– 'Early Rise-Rapid Decline (ER-RD),' 'Medium Rise-Slow Decline (MR-SD),' and 'Delayed Rise-No Decline (DR-ND).'

Table 6: **The ER-SD and DR-ND clusters are aligned with identical trajectories in literature**

<table>
<tr><th>Final clusters</th><th>Brief description</th><th>Identical clusters in prior literature</th></tr>
<tr><td>

Early Rise-
Slow Decline

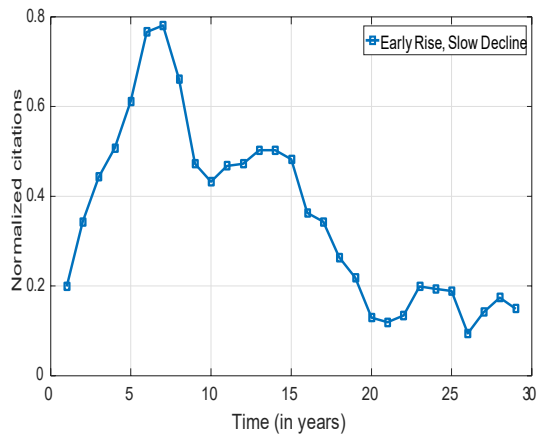

</td><td>

**Long-term trajectory**

- *Growth time:* 6 years
- *Decay time:* 20 years
- *No. of papers:* 17,479

**Short-term trajectory**

- *Growth time:* 5 years
- *Decay time:* 3 to 4 years
- *No. of papers:* 87,956

</td><td>

- Delayed rise, slow decline [13]
- Discovery papers & hot papers [16]
- Normal documents [17]
- Medium rise, slow decline [14]
- Middle-of-the-roads [26]
- Sticky-knowledge-claims [11]
- PeakInit [23]
- Normal-low and Normal-high [27]

</td></tr>
<tr><td>

Delayed Rise-
No Decline

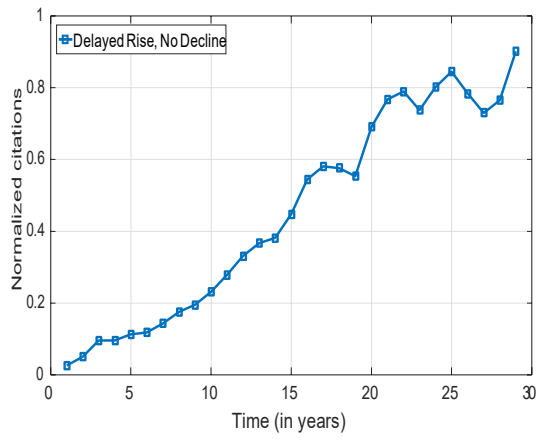

</td><td>

**Long-term trajectory**

- *Growth time:* 29 years
- *Decay time:* no decay
- *No. of papers:* 23,917

**Short-term trajectory**

- *Growth time:* 9 years
- *Decay time:* no decay
- *No. of papers:* 1,03,373

</td><td>

- Delayed rise [17]
- Delayed rise, no decline [14]
- Marathoners [26]
- MonInc [23]
- Delayed recognition papers [28]
- Evergreens [27]

</td></tr>
</table>

The ER-RD and MR-SD clusters have a growth time of 2 to 3 years and 4 to 5 years, respectively. The exact decay time is not mentioned. DR-ND are defined as papers with a slow citation rise initially and receiving a significant proportion of citations only after 12 years. No decay is observed. The ER-RD, MR-SD, and DR-ND clusters are identical to this study's *ER-RD*, *ER-SD*, and *DR-ND* clusters.

Costas et al. [17] identified 3 clusters– 'Flashes-in-the-Pan (FP),' 'Normal Documents (ND),' and 'Delayed Documents (DD).' FP is defined as papers with a growth time of 3 years; however, they are not cited in the long term. ND is defined as papers with a 4 to 5 years growth time followed by an exponential decay. The exact decay times are not mentioned. DD is defined as papers with a growth time of 10 to 11 years and receiving a significant proportion of their citations later than normal documents. They receive citations even after 20 years. The FP, ND, and DD clusters are identical to this study's *ER-RD*, *ER-SD*, and *DR-SD* clusters. Moreover, S. Redner [16] identified three clusters– Sleeping Beauties (SB), Discovery Papers (DP), and Hot Papers (HP), respectively. The behavior of DP and HP align with the *ER-SD* cluster. Their growth time is 4 to 6 years. Besides, SBs align with the *DR-ND* and receive citations 40 years after publication.

S. Baumgartner and L. Leydesdorff [11] identified two clusters– Transient-Knowledge-Claim (TKC) and Sticky-Knowledge-Claim (SKC). Papers belonging to TKC show a typical early peak in citations followed by a steep decline. Papers belonging to SKC have a growth time of 3 to 4 years, and they continue to be cited even after more than 10 years. The TKC and SKC are identical to this paper's *ER-RD* and *ER-SD* clusters, respectively.

Chakraborty et al. [23] defined six clusters– PeakInit, MonDec, MonInc, PeakLate, PeakMult, and Others. PeakInit papers have a growth time of exactly 5 years. MonDec papers have a growth time of 1 year followed by a monotonic decrease. MonInc papers have a growth time of 20 years and no decay. PeakLate papers have a growth time of $>$ 5 years. The exact decay times are not determined. 45% of the papers fell into the Others category, whose trajectory characteristics are not defined. The PeakInit, MonDec, MonInc, and PeakLate clusters are identical to *ER-SD*, *ER-RD*, *DR-ND*, and *DR-SD* from this study. Gou et al. [34] also identified the PeakMult cluster and studied it as a 'literature revival' phenomenon. All-Element-Sleeping-Beauties, a sub-category of SBs, are also a PeakMult cluster. However, we observed that papers of all trajectories receive multiple peaks. The only difference is the time of occurrence of peaks and their varying intensities in individual trajectories. For instance, *ER-SD* received a maximum number of peaks during the growth period, and DR-ND received them during the decay period. Thus, it is one of the inherent properties of a trajectory.

G. Colavizza and M. Franceschet [26] identified 3 clusters– sprinters, middle-of-the-roads, and marathoners. The exact growth and decay times are not explicitly mentioned. Sprinters are defined as papers with fast and high peak values followed by equally rapid aging. Middle-of-the-roads are defined as papers that attain fast but moderate peaks with a gradual decay over time. Marathoners are defined as papers that start slow, peak moderately, keep receiving a higher proportion of citations over a prolonged time, and, finally, citations decline slowly. The sprinters, middle-of-the-roads, and marathoners are identical with *ER-RD, ER-SD,* and *DR-SD* clusters of this study.

Zhang et al. [27] identified four clusters– normal low, normal high, delayed documents, and evergreens. Papers belonging to normal low and normal high clusters receive an early peak followed by a slow decline. Delayed documents are papers with a slow citation rise followed by a slow decay. Evergreens are papers with a continual increase in citations and no decay in the 30 years analyzed. The normal-low and normal-high clusters are similar to *ER-SD,* delayed documents are identical to *DR-SD,* and evergreens align with *DR-ND* clusters of this study. F. Ye and L. Bornmann [29] identified two clusters– Smart Girls (SG) and Sleeping Beauties (SB). SGs are papers with a growth time of 5 years, and the citation angle is $> 60^o$. SBs are papers with a $> 5$ years growth time, and the citation angle is $> 30^o$. SBs receive a major citation proportion after 15 years. The SGs and SBs are identical with *ER-RD* and *DR-SD* of this study.

Bornmann [28] identified two clusters– Hot Papers (HP) and Delayed Recognition (DR). The HP and DR are identical to *ER-RD* and *DR-SD* clusters from this study. Besides, many works only study an extreme trajectory, that is, sleeping beauties [15, 31, 18, 20, 30]. Such papers receive negligible citations for a long time after publication and then, depending upon the awakening intensity, suddenly jump to receive large citations. The decay time is characterized by lower annual citations than peak [28]. It aligns with the *DR-SD* cluster of this study.

To summarize, the qualitative comparison validates our proposed methodology. We can detect all probable classes of trajectories identified in the existing literature. Table (5, 6) shows that we can universally categorize trajectories into four clusters– *ER-RD, ER-SD, DR-ND,* and *DR-SD*. Papers with an *ER-RD* trajectory have a growth period of 3 years and decay in the next 2 years. Short-term trajectories mostly exhibit such a pattern. They receive $n_{pl}$ peaks during the growth period. Besides, both short-term and long-term trajectories exhibit *ER-SD* patterns. Compared to the other two clusters, short-term trajectories with an *ER-SD* pattern receive maximum citations. They have a growth period of 5 years and decay in the next 3 to 4 years. Further, long-term citation trajectories with an *ER-SD* pattern have a growth period of 6 years and decay in the next 20 years. Compared to all other clusters, they receive a maximum $n_{Il}$ and $n_{Im}$ peaks during the growth period.

Both short-term and long-term trajectories also exhibit *DR-ND* patterns. Short-term trajectories with a *DR-ND* pattern have a delayed growth period of 9 years. Further, long-term citation trajectories with a *DR-ND* pattern have a growth period of 29 years and receive maximum citations compared to the other two clusters. No citation decline is seen for the period analyzed. Finally, papers with a *DR-SD* trajectory have a growth period of 16 years and decay in the next 14 years. Long-term trajectories mostly exhibit such a pattern. They receive a maximum $n_{Il}$ peaks during decay.

## 1. Conclusion

The study of clustering citation trajectories reveals that not all articles receive immediate success after publication. A detailed review of prior literature reveals that identical trajectories were detected as different clusters. It is due to the use of arbitrary thresholds and parameter-dependent methods. Besides, different methods captured different groups of trajectories leading to ambiguities in their specific number.

This study proposes a feature-based multiple k-means cluster ensemble framework. It is a generalized framework for capturing the temporal evolution of any trajectory using a generic feature set and clustering them using the ensemble learning method. Four distinct trajectories are obtained- *ER-RD, ER-SD, DR-ND,* and *DR-SD*. A comprehensive comparison reveals that these four clusters can define all prior groups of trajectories identified in the literature. Thus, the issue of ambiguities regarding their distinct number is resolved. Most papers fall into *ER-SD* and *DR-ND* clusters. A negligible share of articles fall into *ER-RD* (widely studied as hot papers) and *DR-SD* (widely studied as sleeping beauties). Further, the *ER-RD* is a characteristic of short-term trajectories, and *DR-SD* is a characteristic of mostly long-term trajectories. Delayed-rise papers receive higher total citations than early risers as they receive citations for a more extended period. However, multiple peaks are detected highest for early risers, establishing that the citations' intensity is higher for them. Self-citations are not excluded from our analyses which is a limitation of this study. Future studies could use other empirical data sets to examine the effectiveness of this methodology.